# Electrical Modulation of Superconducting Critical Temperature in Liquid-Gated Thin Niobium Films

Jiman Choi[1,2,a], Pradheesh R.[1,a], Hyungsang Kim[2], Hyunsik Im[2], Yonuk Chong[1,b], and Dong-Hun Chae[1,c]

[1] *Korea Research Institute of Standards and Science, Daejeon 305-340, Korea*
[2] *Division of Physics and Semiconductor Science, Dongguk University, Seoul 100-715, Korea*

**Abstract**

We demonstrate that the superconducting critical temperature ($T_c$) of thin niobium films can be electrically modulated in a liquid-gated geometry device. $T_c$ can be suppressed and enhanced by applying positive and negative gate voltage, respectively, in a reversible manner within a range of about 0.1 K. At a fixed temperature below $T_c$, we observed that the superconducting critical current can be modulated by gate voltage. This result suggests a possibility of an electrically-controlled switching device operating at or above liquid helium temperature, where superconductivity can be turned on or off solely by the applied gate voltage.

[a] These authors contributed equally to this work.
[b] Electronic address: yonuk@kriss.re.kr
[c] Electronic address: dhchae@kriss.re.kr

Electric field control of superconductivity has been a research topic for more than five decades since Glover and Sherrill observed a signature of critical temperature change in superconducting indium and tin[1]. In nonmetallic systems with intrinsically low carrier density, electrical tunability of superconductivity has been demonstrated in the dielectric field effect geometry[2,3]. Unlike low carrier density systems such as semiconductors, modulation of carrier density and conductivity in metals is challenging due to the effective screening by the conduction electrons[4,5].

The field effect technique incorporating ionic media has been developed since 1950s[6,7], and it has been recently revived to overcome the limits in carrier modulation capability for solid state research[8-17]. At the interface between an ionic medium and a solid surface, ions in the medium and induced surface carriers form an electrical double layer with a nanometer scale separation under applied gate voltage. The electric field at the interface can reach as high as 100 MV/$cm^2$ and it leads to high surface carrier density up to $10^{15}$/$cm^2$ [9,16,17]. Recent experiments showed that correlated states of matter such as superconductivity[8,10,11,13] or ferromagnetism[12] can be induced by the ionic-liquid gating in nonmetallic materials. The following application of this technique to metallic systems showed that modulation of carrier density and conductance can be orders of magnitudes larger than the conventional field effect geometry with dielectric insulators[14-16].

In this Letter, we report electrical modulation of electronic properties of a niobium thin film, which is the most widely used material for superconducting electronics. We have observed that the superconducting critical temperature ($T_c$) of niobium thin films can be electrically modulated in a liquid-gated geometry. The critical temperature change depends on the polarity and magnitude of the applied gate voltage; $T_c$ is enhanced (suppressed) with negative (positive) gate voltage. The tunable critical temperature range is close to 0.1 K and this value is three orders of magnitudes larger than the previously reported values in superconducting films[1].

The devices are fabricated in Hall bar geometry by optical lithography and dry etching process as shown in Fig. 1(a) and 1(b). The niobium film was deposited on a c-plane (0001) sapphire substrate by dc magnetron sputtering. In this experiment, we chose the sapphire substrate due to its superior insulating property. In general, niobium film on sapphire substrate is also of interest since sapphire has excellent high frequency properties at cryogenic temperatures. The film deposition condition was tuned such that the film stress is slightly compressive but less than 200 MPa. The superconducting coherence length of the Nb film deduced from $H_{c2}$ measurement and Werthamer-Helfand-Hohenberg analysis[18] is 15 nm at 4 K. In films with thicknesses less than the coherence length, the critical temperature is suppressed as shown in Fig. 1(c). This enables us to systematically tune the critical temperature by changing the film thickness. For this study, we used an 8 nm niobium film with nearly 4.2 K transition temperature. A Hall bar pattern shown in Fig. 1(a) was defined by standard photolithography and Ar:SF$_6$ reactive ion etching. After the plasma etching, the residual photoresist was carefully removed with warm organic solvents with the sequence of N-methyl pyrollidinone, acetone, isopropyl alcohol rinsing and nitrogen blow-dry. In a glove box under dry argon atmosphere, adequate amount of ionic liquid, N,N-diethyl-N-(2-methoxyethyl)-N-methylammonium bis-(trifluoromethylsulfonyl)-imide (DEME-TFSI) was dropped to cover the Hall bar channel and a large coplanar gate electrode. Figure 1(b) shows a typical device with ionic liquid. The active channel size of this device is 20 μm long and 10 μm wide. The leakage current from the gate electrode to the niobium channel was less than 1 nA. The device was mounted in a temperature-controlled cryostat, and the sample space was evacuated down to $10^{-5}$ Torr with a cryopump during measurement. Electrical measurement was performed with the lock-in technique in the four probe configuration. The gate voltage was always applied at 230 K, and then the temperature was changed for measurement.

Figure 2 shows gate voltage dependence of the resistance of an 8 nm thin Nb film measured at 230 K. This setting temperature is higher than the glass transition temperature (190 K) of the DEME-TFSI for ionic mobility[9]. The reversible change of the measured resistance is as large as 0.05 % from 0 to 5V gate voltage sweep. We note that irreversible change of resistance was observed in the negative gate voltage unlike the positive gate voltage scan in Fig. 2. (Here the polarity is defined as the potential of the gate electrode relative to the measured niobium channel.) We speculate this irreversible change to the anodization process of niobium under electrical bias. In practical purpose, this irreversibility does not affect the modulation of the superconducting properties as described below. However, if this irreversibility needs to be avoided, we can apply gate voltage to the device only in positive polarity.

Figure 3(a) shows that the superconducting transition temperature is tuned with applied gate voltage in a reversible manner. This is the key observation in this work. The temperature was changed slowly near the superconducting transition at the rate of 0.2 K per minute to avoid apparent hysteresis between cool-down and warm-up. The hysteresis in temperature sweep is less than 5 mK. In Fig. 3(a), only data sets acquired with descending temperature are shown for convenience. The measurement sequence regarding applied gate voltage was 0 V, 5 V, 3 V, 0 V, -4 V, -2 V, and finally back to 0 V. Every time we change the gate voltage, the device temperature was raised up to 230 K and then cooled down again with a new value of applied gate voltage. Although we observed irreversible change of resistance in negative gate voltages, a good consistency of the superconducting transition curves at zero gate voltage for the initial and final measurements indicates that the modulation of the critical temperature does not arise from an irreversible chemical process. Figure 3(b) summarizes the gate voltage dependence of $T_c$. Here $T_c$ was defined as the temperature with the maximum differential resistance with respect to temperature. Depending on the gate voltage polarity, $T_c$ is enhanced with negative voltages while it is suppressed with positive voltages. The tunable temperature range was about 80 mK. We observed a qualitatively similar behavior from two other devices having the same thickness. We note that this change is almost three orders of magnitudes larger than the values observed in Ref. 1

In the superconducting state far below the critical temperature, we also observed that the critical current can be changed by gate voltage. Figure 4 shows the current-voltage characteristics of another niobium thin film for three different gate voltages at 1.9 K ($T_c$ = 4.2 K). This clearly demonstrates that the superconducting critical current can be electrically modulated in thin niobium films. We observed the critical current modulation of 18 µA out of the full critical current of 208 µA in a 10 µm wide and 8 nm thick niobium strip.

In order to utilize this effect in a practical device, the superconducting critical current at each applied gate voltage is also an important factor. Figure 5 shows the measured $T_c$ as a function of bias current for three gate voltages. For a given gate voltage, the superconducting transition was monitored by varying bias current from 0.5 µA to 50 µA. From Fig. 5, we can deduce the critical current as a function of temperature at different gate voltages. We can see that the typical critical current modulation at a fixed temperature in this device is about 20 µA. As an example of a switching device, if we focus on the case of 4.25 K, switching between the superconducting and non-superconducting states would be possible with the maximal supercurrent in the superconducting state as large as 20 μA. Since most of the superconducting electronics based on niobium is operated at the liquid helium temperature(4.2 K), this is a clear demonstration for a possible switching device operating at practical temperature.

In order to understand the origin of electrical modulation of superconducting transition, we also performed experiments with a thicker niobium film as thick as 120 nm. We have observed very similar modulation of the superconducting properties; both suppression and enhancement of $T_c$ were observed in this thick device. We should note that the electrostatic field can lead to the carrier density modulation only at the surface of metal film, not in the inner (bulk) part. Therefore, if $T_c$ is modulated by the carrier density change at the surface, suppression of $T_c$ cannot be explained by the simple electrostatic effect on metal surface. This makes it difficult to simply interpret our observations as a pure electric field effect. We thus cannot rule out extrinsic effects such as electro-mechanical effect due to the electrostriction of the polarized ionic liquid in the solid phase at the interface. Further study is required to understand the origin of the observed modulation of superconductivity.

In summary, we have observed electrical modulation of the resistance, the superconducting critical temperature, and the critical current in thin niobium films with a liquid-gated geometry. Although the exact origin of this electrical modulation of superconducting transition is not clear yet, this effect opens up a possibility of an electrically switchable device operating at or above liquid helium temperature. The practicality of device comes from the following facts; 1) it is based on the most commonly used superconducting material of niobium, 2) it shows potential supercurrent handling as large as 20 μA, and 3) the critical temperature can be electrically modulated across the liquid helium temperature.

This work was supported by the Creative Research Program (KRISS-2014-14011031 and KRISS-2014-14011025) of Korea Research Institute of Standards and Science. D. C. also thanks F. Paolucci and J. Smet in Max Planck Institute for Solid State Research, Stuttgart for motivating discussions in the initial stage of the ionic gating project.

# Figure Captions

FIG. 1. (a) A typical Hall bar device of niobium film. Bright region is niobium film and dark region is sapphire substrate. (b) Device with ionic liquid. (c) Superconducting transition temperature *vs*. niobium film thickness. The dashed line is only guide to the eyes. (4) Schematic diagram of ionic liquid-gating geometry.

FIG. 2. Gate voltage( $V_g$ ) dependence of resistance in 8 nm niobium film. The temperature was 230 K and the gate voltage sweep speed was 1 mV/s. Inset depicts the four probe measurement configuration.

FIG. 3. (a) Gate voltage dependence of superconducting transitions. The measurement sequence regarding gate voltage was 0 V→ 5 V →3 V→ 0 V→ -4 V→ -2 V → 0 V. The bias current was 0.5 µA. (b) Superconducting critical temperature, $T_c$ *vs*. gate voltage, $V_g$. $T_c$ was extracted as the temperature with maximum differential resistance with respect to temperature.

FIG. 4. Current-voltage characteristics at 1.9 K measured for different gate voltages. This data was taken in a different device from Figs. 2, 3 and 5 but with the same film thickness and device geometry. All 8 nm thick devices have the superconducting transition temperature close to 4.2 K.

FIG. 5. Superconducting critical temperature *vs*. bias current at different gate voltages.

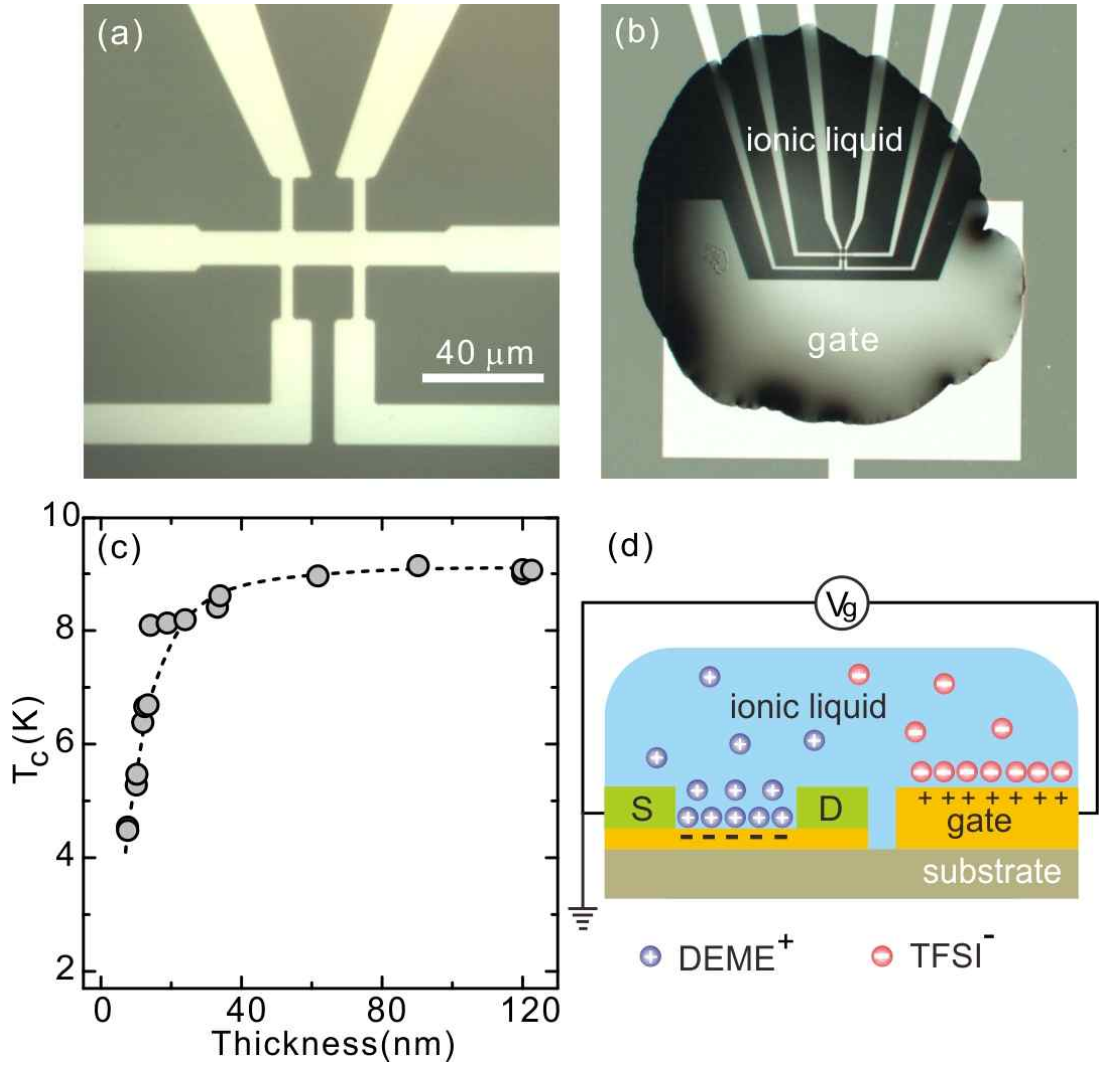

**FIG. 1**

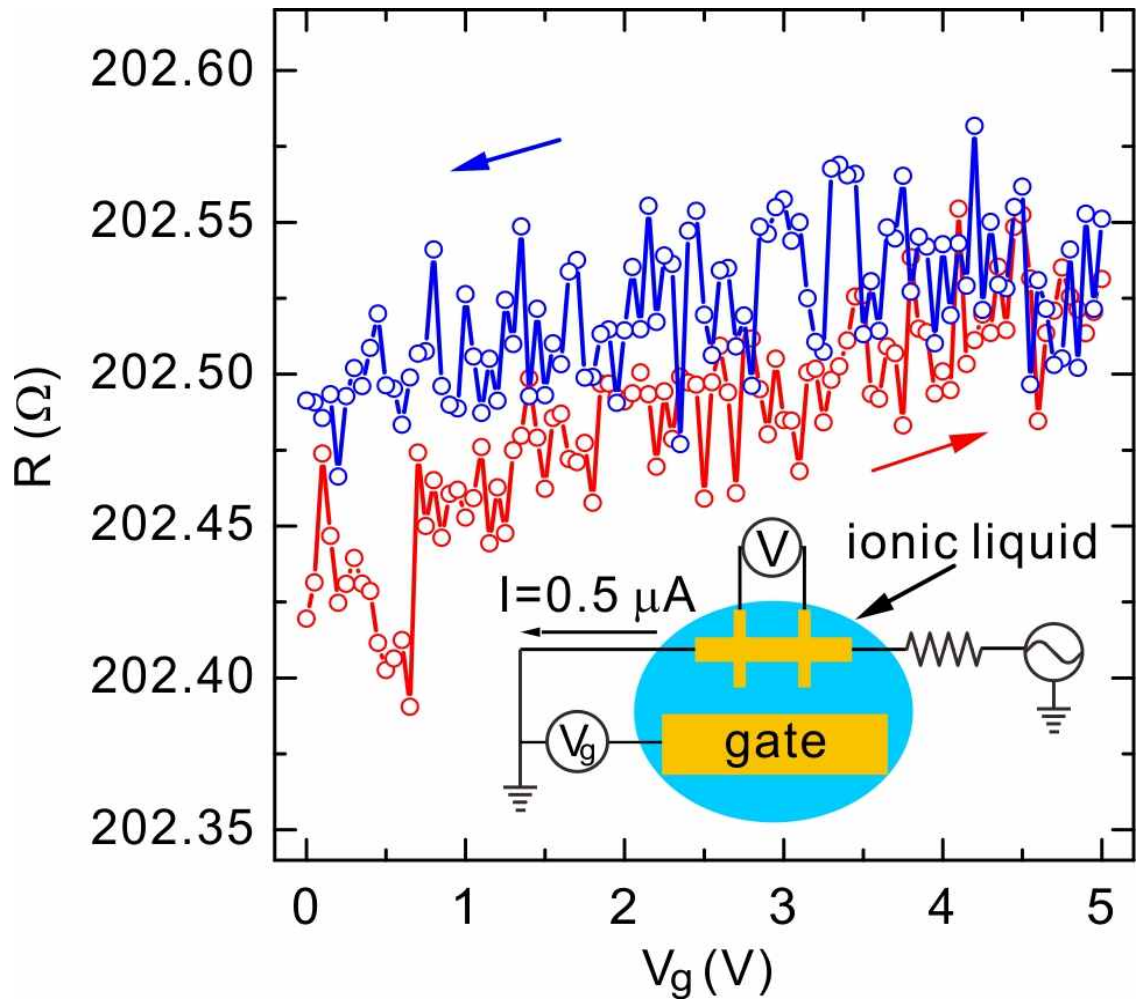

**FIG. 2**

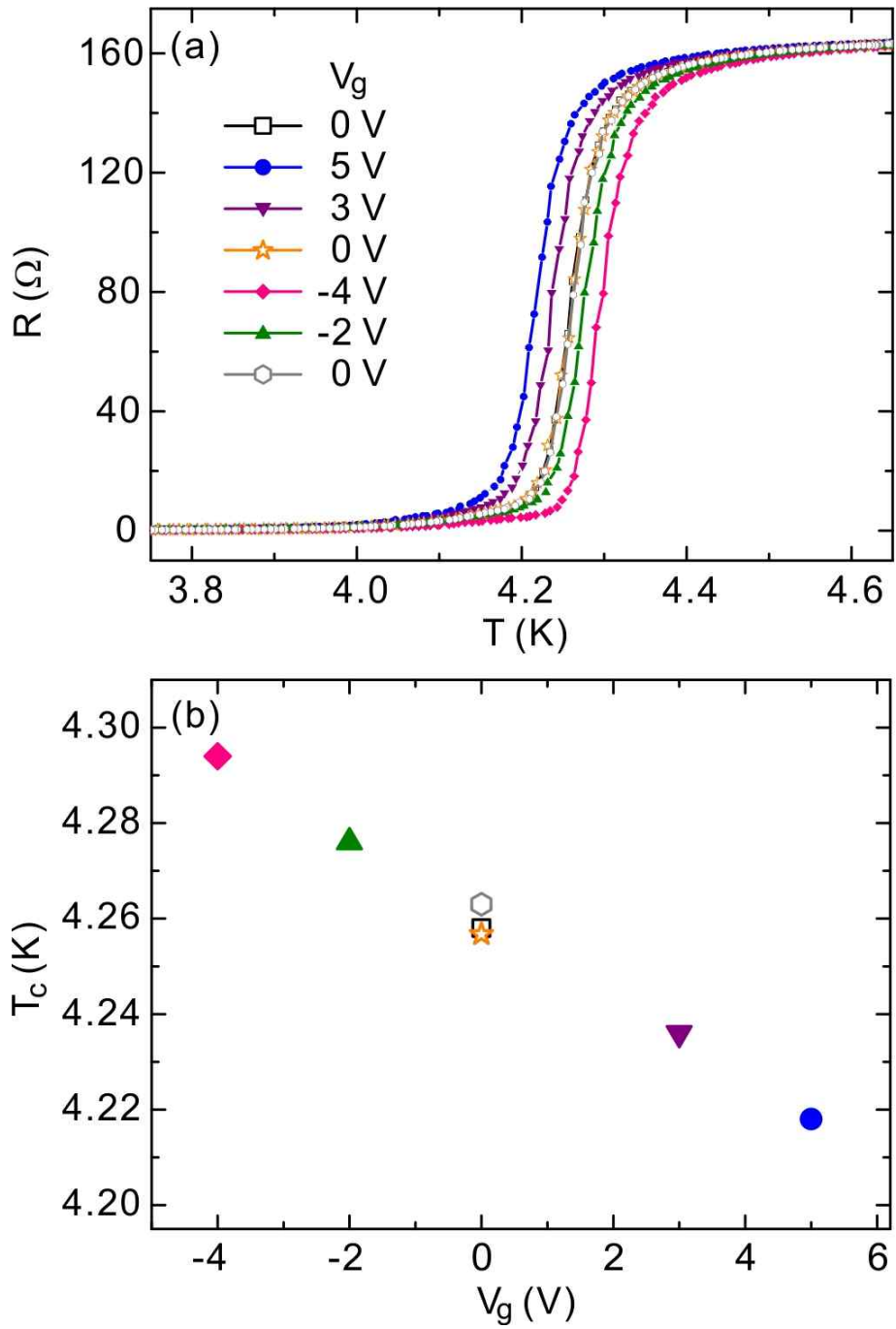

**FIG. 3**

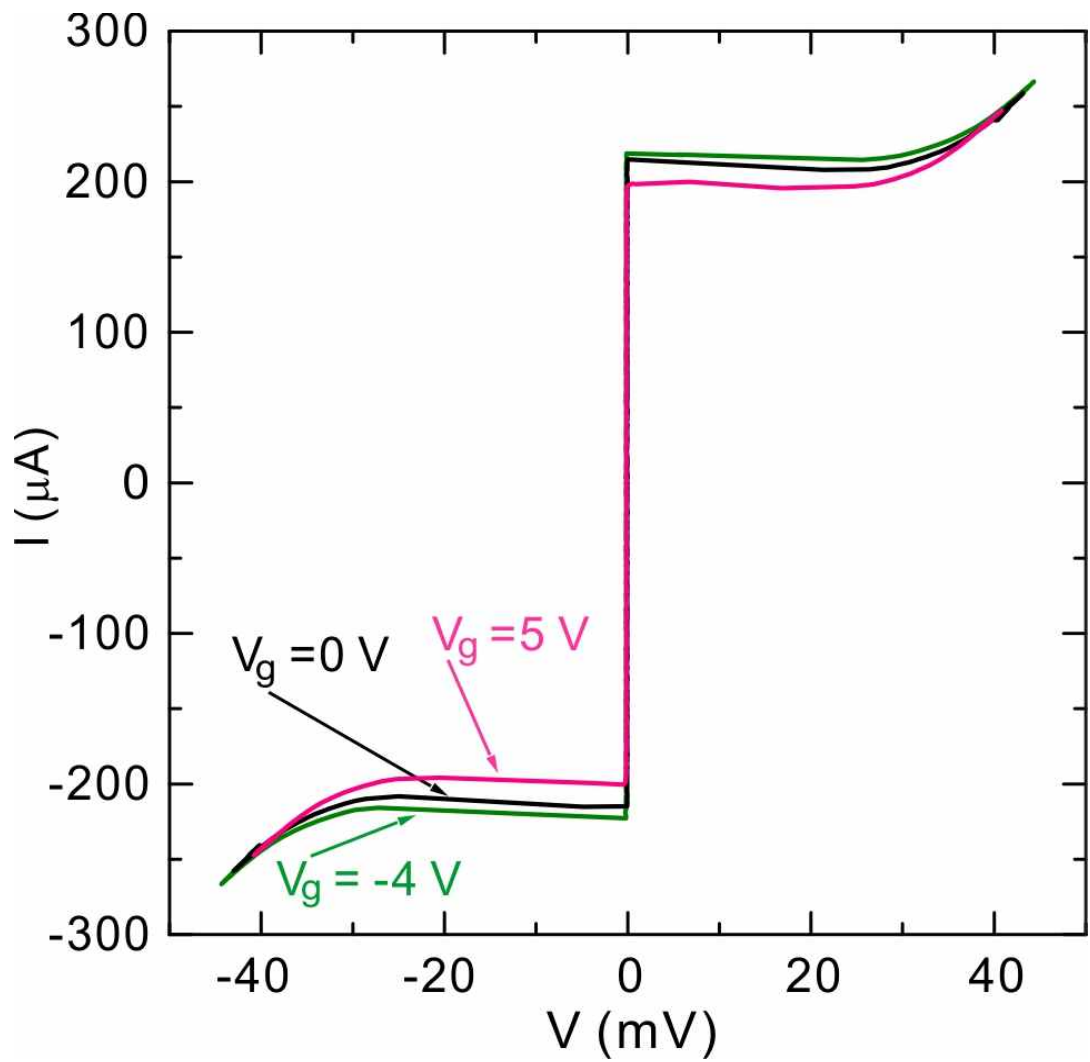

**FIG. 4**

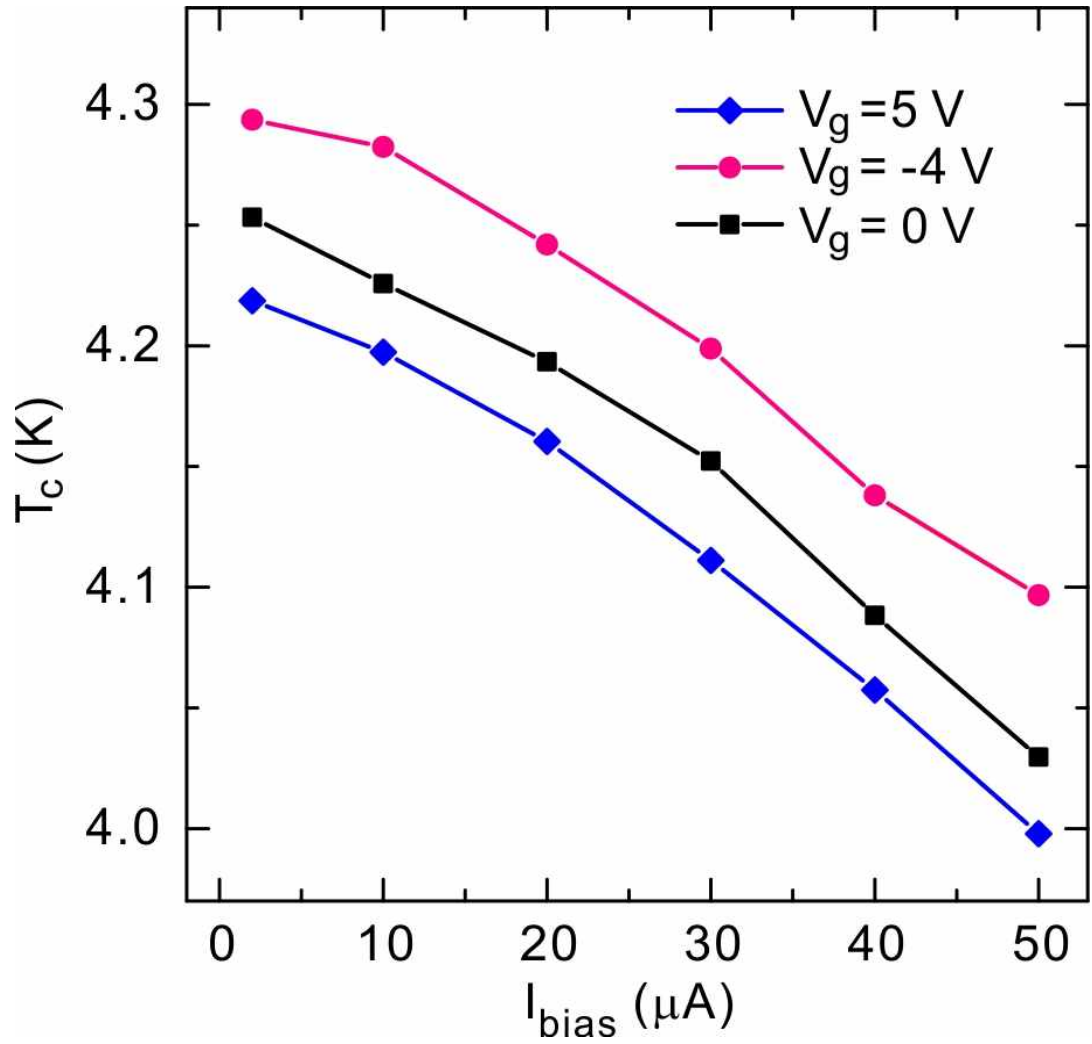

**FIG. 5**